\documentclass[lettersize,journal]{IEEEtran}
\usepackage{amsmath,amsfonts}
\usepackage{algorithmic}
\usepackage{algorithm}
\usepackage{array}
\usepackage[caption=false,font=normalsize,labelfont=sf,textfont=sf]{subfig}
\usepackage{textcomp}
\usepackage{stfloats}
\usepackage{url}
\usepackage{verbatim}
\usepackage{graphicx}
\usepackage{cite}
\usepackage{xcolor}
\newcommand*{\corr}[1]{\textcolor{black}{#1}}

\begin{document}

\title{A 4–20 GHz Superconducting 90$^\circ$ Hybrid Using Capacitively  Loaded Coupled Lines for 2SB Heterodyne Receivers}

\author{Fran\c{c}ois Joint,  Max Behrens, Denis Meledin, Cristian Lopez,  Magnus Strandberg, Sven-Erik Ferm, Mathias Fredrixon, Leif Helldner, Alexei Pavolotsky, Erik Sundin, Igor Lapkin, Victor Belitsky, Vincent Desmaris
\thanks{F. Joint, M. Behrens, D. Meledin,  A. Pavolotsky, M. Strandberg, S.-E. .Ferm, M. Fredrixon, L. Helldner, E. Sundin, I. Lapkin, V. Belitsky, V. Desmaris are with the Department of Physics and Astronomy 
Chalmers University of Technology, Onsala Space Observatory,SE-439 92 Onsala, Sweden. C. Lopez is now with Maxwell Applied Technologies, Televes Corporation, Spain.
}
}
\maketitle

\begin{abstract}
We report the design and cryogenic characterisation of a broadband superconducting 90$^\circ$ intermediate-frequency (IF) hybrid covering 4–20 GHz for sideband‑separating (2SB) heterodyne receivers. The hybrid is implemented in thin-film niobium using capacitively loaded edge coupled transmission lines, avoiding the air-bridge interconnections required in Lange-coupler layouts while retaining a compact superconducting implementation with low loss and good quadrature balance. Cryogenic measurements using Vector Network Analyser (VNA) show amplitude and phase imbalance better than 0.6 dB and 6$^\circ$, respectively, across the full band. When used in a 2SB receiver, this level of IF balance is consistent with the observed image-rejection performance in the 10–20 dB range, for the full receiver. The design supports the development of wide-instantaneous-bandwidth 2SB receivers for millimeter and submillimeter observatories, improving spectral grasp for broadband line surveys and high-redshift line searches.
\end{abstract}


\section{INTRODUCTION}
\IEEEPARstart{S}{ideband} seperating (2SB) heterodyne receivers are now the preferred architecture for millimeter/sub-‐ millimeter spectral‑line astronomy because they reject atmospheric and instrumental noise that folds in from the image band in double‑sideband systems \cite{belitsky_sepia_2018, meledin_sepia345_2022, vassilev_swedish_2008, belitsky_alma_2018, mahieu_alma_2012, sekimoto2008development, khudchenko_design_2019, baryshev_alma_2015, uzawa_development_2013}. In a 2SB front end, the RF signal is split into two equal‑amplitude branches in quadrature, both mixers are pumped in phase, and the resulting IF signals are recombined in a 90$^\circ$ IF hybrid so that the upper and lower sidebands emerge at separate output ports. This architecture is used in ALMA receivers and remains central to receiver developments associated with the ALMA Wideband Sensitivity Upgrade (WSU), where accurate spectral-line observations require high image rejection over broader IF bandwidths~\cite{carpenter_alma2030_2023}. \corr{While the original ALMA 2SB receiver specification required an image rejection ratio (IRR) of at least 10~dB, current ALMA system-level requirements call for more than 13~dB over the full applicable IF range and more than 15~dB over 90$\%$ of that range\cite{ALMA2024_80_04}.}

In practice, the achievable image-rejection ratio (IRR) is determined by the cumulative amplitude and phase imbalance of the complete 2SB signal path, including the RF quadrature network, the mixer pair, and the IF quadrature hybrid. Even modest imbalances cause imperfect cancellation of the unwanted sideband and degrade the effective single-sideband noise temperature through finite image rejection. \corr{Depending on the receiver architecture, the IF hybrid may be placed before or after the first cryogenic low-noise amplifiers. When located ahead of the first amplifier, its insertion loss contributes directly to the receiver noise temperature, requiring both low cryogenic loss and tight amplitude and phase balance over the full IF bandwidth. Alternatively, the mixer outputs may be amplified before recombination in the IF hybrid. In this configuration, the hybrid loss is largely suppressed by the preceding amplifier gain, but the input match of the amplifiers and differential gain and phase between the two IF chains become additional contributors to the receiver noise and image-rejection performance. Reflections between the mixers and amplifiers may also perturb the effective mixer embedding impedance and further degrade the balance between the two signal paths.} Prior superconducting IF hybrids for ALMA‑class receivers based on Nb Lange‑coupler implementations with integrated bias‑T \cite{rashid_superconducting_2014, billade_superconducting_2012} demonstrated the viability of cryogenic thin film hybrid integration and quantified how IF amplitude and phase balance affect system IRR; Separately, cryogenic 90$^{\circ}$ coaxial hybrids have shown that sub‑dB amplitude and near‑degree phase balance are achievable at 20 K when microwave components are engineered explicitly for cold operation \cite{kooi_multioctave_2023, malo-gomez_cryogenic_2009, lopez2023description}.
\begin{figure}[!t]
\centering
\includegraphics[width=3in]{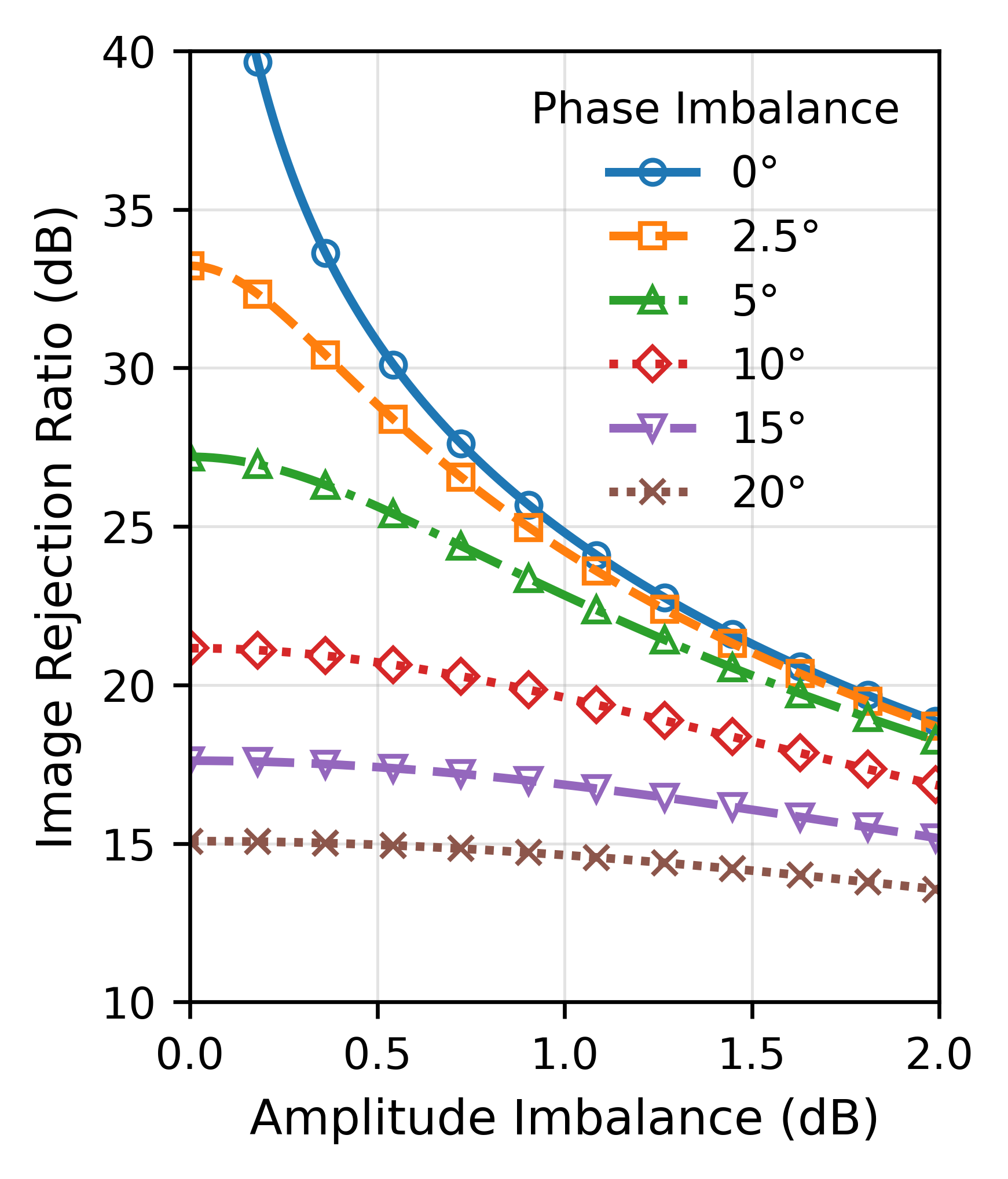}
\caption{Calculated IRR as a function of amplitude imbalance for several fixed phase-imbalance values. The curves illustrate the sensitivity of image rejection to combined amplitude and phase errors. In this work, $\leq 0.5$~dB amplitude imbalance and $\leq 5^\circ$ phase imbalance are used as IF-hybrid design targets within a receiver-level imbalance budget aimed at an IRR of order 25~dB.}
\label{fig:srr}
\end{figure}
At the same time, modern backends are pushing for much wider instantaneous IF. The ALMA Wideband Sensitivity Upgrade \cite{carpenter_alma2030_2023} explicitly targets octave‑to‑multi‑octave IF signal chains, up to 4–20 GHz per sideband and polarization, to increase spectral grasp and continuum speed without sacrificing spectral resolution. These system‑level drivers set a challenging requirement for the IF hybrid: maintain low insertion loss with sub‑dB/degree balance across $\geq$2 octaves, operating at cryogenic temperature and within tight form‑factor and thermal budgets.

This work addresses that need. We report a superconducting 4–20~GHz quadrature IF hybrid engineered for 2SB receiver. Unlike prior superconducting IF hybrids based on Lange couplers, our device uses edge coupled \corr{microstrip} transmission lines periodically loaded with Metal-Insulator-Metal (MIM) coupling capacitors to synthesize the required 3~dB quadrature coupling over more than two octaves, in a layout compatible with cryogenic assembly. Measured at cryogenic temperature, the hybrid exhibits amplitude and phase imbalance better than 0.6 dB and 6$^\circ$, respectively, across 4–20~GHz. \corr{Its system-level performance is also validated in a 2SB receiver operated at an LO frequency of 320~GHz and demonstrate an IRR above 10~dB across the full 4--20~GHz IF band, with average USB and LSB rejection values of approximately 18.5 and 21.9~dB, respectively.} The remainder of the paper is organized as follows. Sections~II and III describe the design implementation and fabrication, respectively. Section~IV presents the cryogenic characterization and discusses system-level considerations for 2SB measurements, followed by the conclusions in Section~V.

\begin{figure}[!t]
\centering
\includegraphics[width=3.4in]{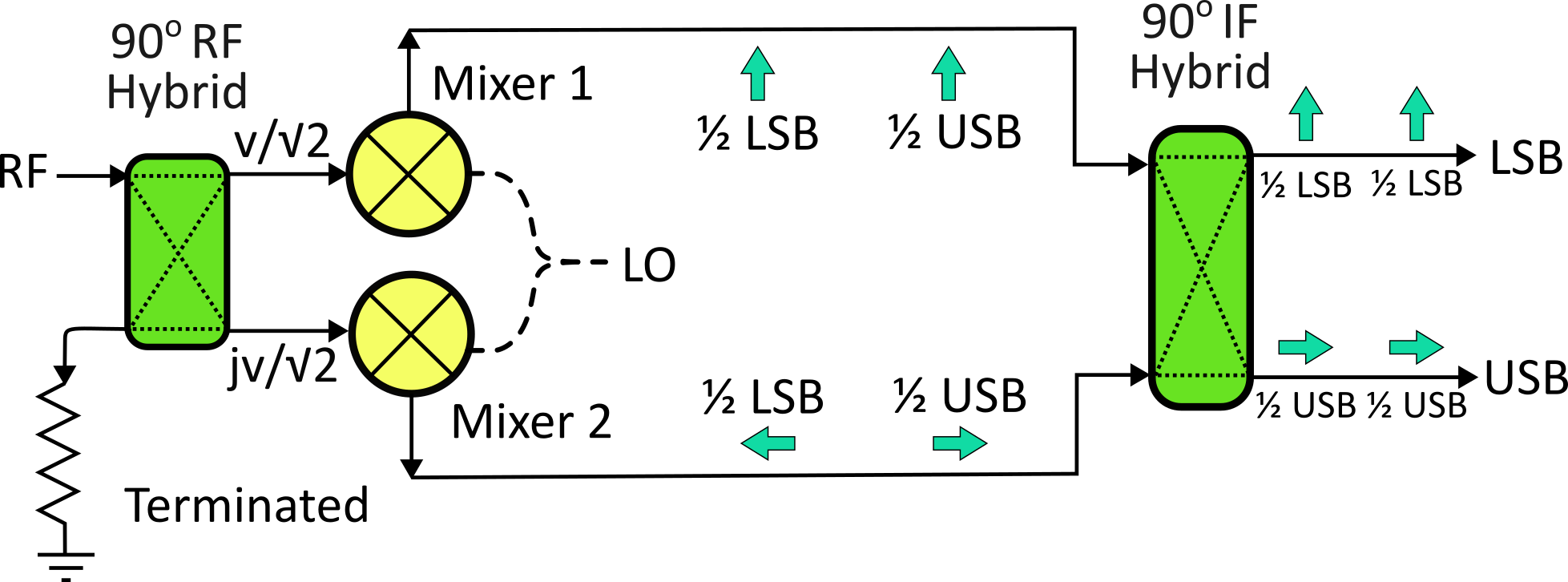}
\caption{ Schematic of a 2SB mixer with quadrature hybrids introducing the appropriate phase shifts for sideband separation.}
\label{fig:2sb_schematic}
\end{figure}

\section{Design}
In a 2SB receiver, the RF signal is split between two nominally identical mixers operated in quadrature. After downconversion, the two IF signals must be recombined such that one sideband adds constructively and the image sideband cancels. This requires equal amplitudes and a relative phase difference of $180^\circ$ for the unwanted sideband at each output. The IF hybrid recombines the two mixer outputs and separates the upper sideband (USB) and lower sideband (LSB) signals, as shown in Fig.~\ref{fig:2sb_schematic}. This requires a low-loss 3-dB quadrature hybrid with small amplitude and phase imbalance over the 4--20~GHz IF band.

The IF Hybrid is directly responsible for recombining the two mixer outputs with the required relative phase shift. Any deviation from ideal amplitude balance or phase shift will allow part of the unwanted sideband to appear at each output. This residual sideband acts as an additional noise contribution to the receiver. The ratio of wanted to unwanted sideband power, also known as the sideband rejection ratio (SRR) or image-rejection ration (IRR), can be related to the system noise temperature. Defining the amplitude imbalance $g$ as the ratio of the signal magnitudes and the phase imbalance $\phi$ as the deviation from the ideal $\pi/2$ phase difference at the hybrid output, the image-rejection ratio can be written in terms of the residual amplitude and phase imbalance \cite{sobis_low_2011,rashid_superconducting_2014, billade_superconducting_2012}:

\begin{equation}
    IRR = \frac{1+g^2+2g\cos(\phi)}{1+g^2-2g\cos(\phi)}
\end{equation}

\corr{For an ideal 2SB receiver with perfectly balanced amplitude and phase, the unwanted sideband cancels completely and the IRR is infinite. In practice, as shown in Fig.~\ref{fig:srr}, the rejection is determined by the accumulated complex amplitude and phase imbalance of the complete receiver. Contributions arise from the RF quadrature hybrid, the relative conversion gain and phase of the two SIS mixers, differential RF and IF electrical paths, the IF hybrid, and impedance mismatches. In particular, parasitic reflections between imperfectly matched components can form asymmetric interference paths and introduce additional frequency-dependent amplitude and phase errors. Such effects have been shown to limit the IRR of 2SB SIS receivers even when the nominal hybrid amplitude and phase balance is already small \cite{khudchenko_modular_2017,hesper_high-performance_2017, khudchenko_design_2023}. These contributions don't create a simple additive error budget, since their relative signs and phases can either add to or partially compensate one another. As a useful reference, an accumulated imbalance of approximately 1.25~dB and 8$^\circ$ corresponds to an IRR of about 20~dB. The IF hybrid should contribute only a fraction of this overall imbalance. In this design, targets of 0.5~dB amplitude imbalance and 5$^\circ$ phase imbalance were adopted for the IF hybrid, which correspond to a rejection of approximately 26~dB when the hybrid is considered in isolation. This gives sufficiently low IF-hybrid imbalance for high-rejection 2SB operation and leaves the final receiver performance to be determined by the combined RF-hybrid, mixer, interconnection, and IF-chain errors. Although this level of balance is compatible with the ideal Chebyshev multi-section synthesis, extending it over the 4--20~GHz bandwidth requires an extremely strongly coupled central section (see Tab.~\ref{tab_0}). The realized capacitively loaded edge-coupled line is therefore a manufacturable approximation to the ideal impedance profile, and the achievable imbalance is ultimately limited by this approximation and by fabrication tolerances.}

\subsection{Five-Section Quadrature Hybrid Synthesis}

Planar quadrature hybrids have been realized using a range of topologies, from branch-line and rat-race couplers to Lange, tandem, and multisection coupled-line implementations. Broadband 3-dB operation is commonly obtained by cascading coupled-line sections, but such designs become sensitive to tight coupling, phase-velocity imbalance, and discontinuity parasitics, which often require compensation \cite{gruszczynski_design_2006}. Recent planar coupler work has therefore explored alternative tight-coupling and loaded-line approaches, including engineered coupled-line geometries for improved directivity \cite{wu_single-layer_2013} and unit cells in which lumped capacitance is used to control the odd-mode response \cite{elsheikh_wideband_2022}. For the present 4--20~GHz IF bandwidth, narrowband topologies such as branch-line \cite{lee_wideband_2012} and rat-race \cite{gruszczynski_broadband_2012} couplers are not suitable. Lange \cite{lange_interdigitated_1969, rashid_superconducting_2014} and tandem \cite{staszek_broadband_2014} realizations become increasingly challenging when strong coupling, phase-velocity control, and fabrication tolerances must be maintained more than two octaves \cite{rashid_design_2016,malo-gomez_cryogenic_2009, khaira_cryogenic_2022}. In particular, a Lange implementation of the center section would require narrow fingers, very small inter-finger spacing, and multiple air bridges; the associated parasitic capacitance and fabrication sensitivity become problematic for the present bandwidth.
\begin{figure}[!t]
\centering
\includegraphics[width=3.4in]{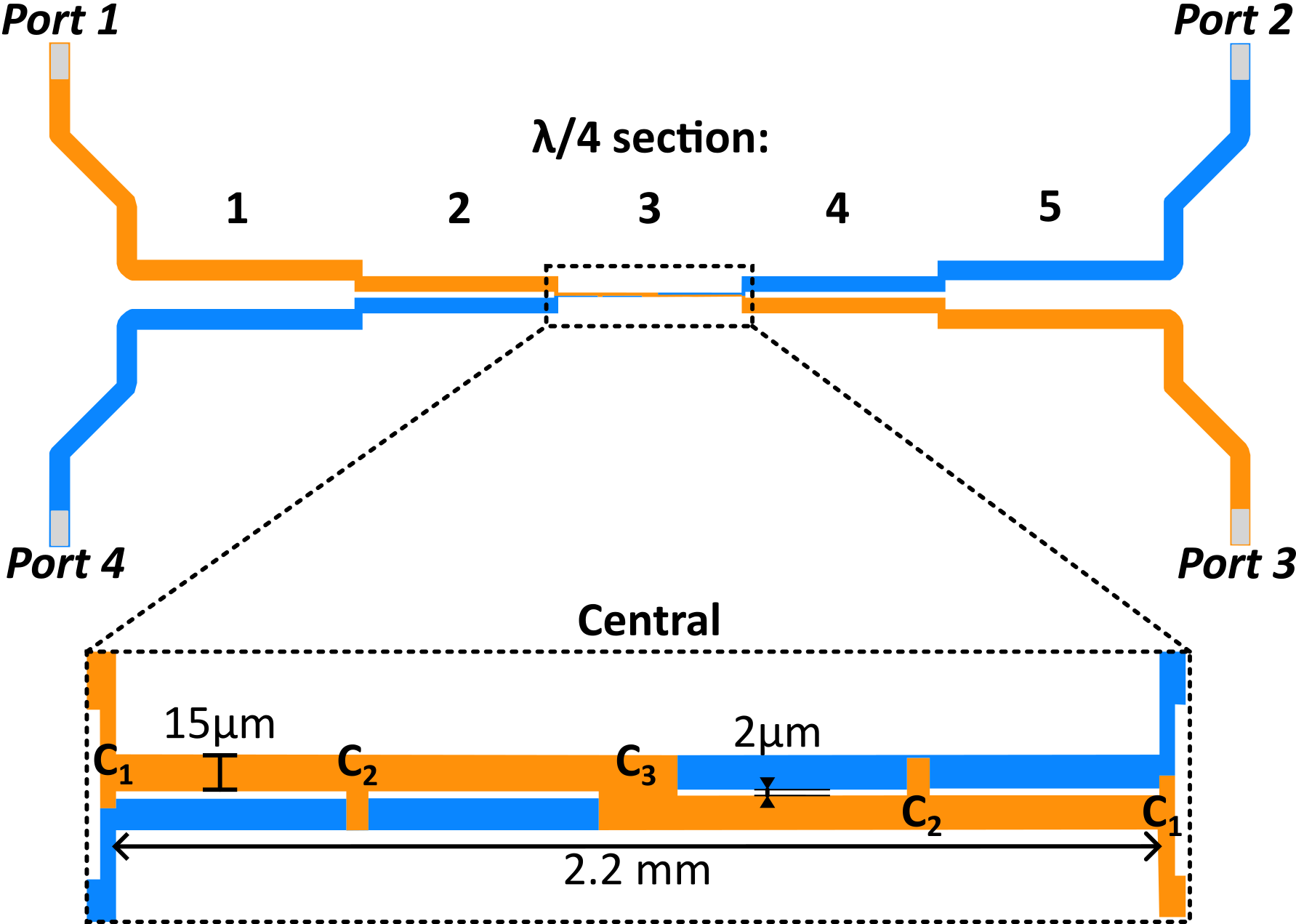}
\caption{Top-view schematic of the five $\lambda/4$ \corr{microstrip} coupled-line sections forming the quadrature hybrid. The weakly coupled sections 1, 2, 4, and 5 are realized as conventional coupled lines, and the central section 3 is divided into four adjacent coupled-line subsections loaded by five MIM capacitors with values $C_1$, $C_2$, and $C_3$. The central capacitor $C_3$ also implements the crossover between the two branches. The grey squares indicate the palladium (Pd) soldering pads used for chip interconnection.}
\label{fig_0}
\end{figure}
Moreover, a normal-metal implementation requires a conductor thickness of several skin depths over the IF band. For gold at cryogenic temperature this leads to electroplated films with thicknesses of order a few micrometers, which is difficult to combine with the small gaps and controlled capacitances required in the strongly coupled center section. In contrast, superconducting niobium provides negligible RF resistance below the gap frequency and allows the hybrid to be implemented using thin-film lithography with sub-micrometer thicknesses. This makes Nb well suited for a compact, low-loss cryogenic IF hybrid.

Based on the above considerations, we adopt a five-section coupled-line quadrature hybrid synthesized using an equal-ripple (Chebyshev) distribution of coupling coefficients. The synthesis is based on the standard even/odd-mode formulation for quarter-wave coupled-line sections from \cite{cristal_theory_1965},
\begin{equation}
Z_{e,i}=Z_0\sqrt{\frac{1+k_i}{1-k_i}}, \qquad
Z_{o,i}=Z_0\sqrt{\frac{1-k_i}{1+k_i}},
\end{equation}
where $k_i$ is the coupling coefficient of the $i$-th section and the design center frequency is taken as $f_0=12$~GHz. The resulting modal impedances are listed in Table~\ref{tab:zezo}.

\begin{table}[t]
\caption{Even- and odd-mode impedances for the 5-section CLIN hybrid.}
\label{tab:zezo}
\centering
\renewcommand{\arraystretch}{1.15}
\setlength{\tabcolsep}{6pt}
\begin{tabular}{c c c}
\hline
\textbf{Section(s)} & $\mathbf{Z_e}$ \textbf{($\Omega$)} & $\mathbf{Z_o}$ \textbf{($\Omega$)} \\
\hline
1 and 5 & 54.73 & 45.68 \\
2 and 4 & 70.45 & 35.49 \\
3 (center) & 205.10 & 12.19 \\
\hline
\end{tabular}
\label{tab_0}
\end{table}

\subsection{Equivalent circuit model of the center section}
Coupled lines are easy to fabricate when the line widths and gaps are reasonably large, as is the case for the weakly coupled sections 1, 2, 4, and 5 (Fig.~\ref{fig_0}). The center section requires a much higher coupling factor and would, in a conventional coupled-line implementation, require an impractically small gap of about 0.1 $\mu m$. To make the design manufacturable, the gap is increased to a value compatible with the fabrication process. The line widths mainly determine the even-mode impedance, and the gap mainly affects the odd-mode impedance. The line widths can therefore be chosen to match the required even-mode impedance, and the gap can be set only by the fabrication constraints. Coupling capacitors are then added to lower the odd-mode impedance and recover the required coupling and impedance for the center section. This process is demonstrated by interleaving manufacturable coupled lines with lumped capacitors:
\[
C_1-\mathrm{CL}_1-C_2-\mathrm{CL}_2-C_3-\mathrm{CL}_3-C_2-\mathrm{CL}_4-C_1
\]
with mirror symmetry about the central capacitor \(C_3\). 

\begin{figure}
\centering
\includegraphics[width=3.4in]{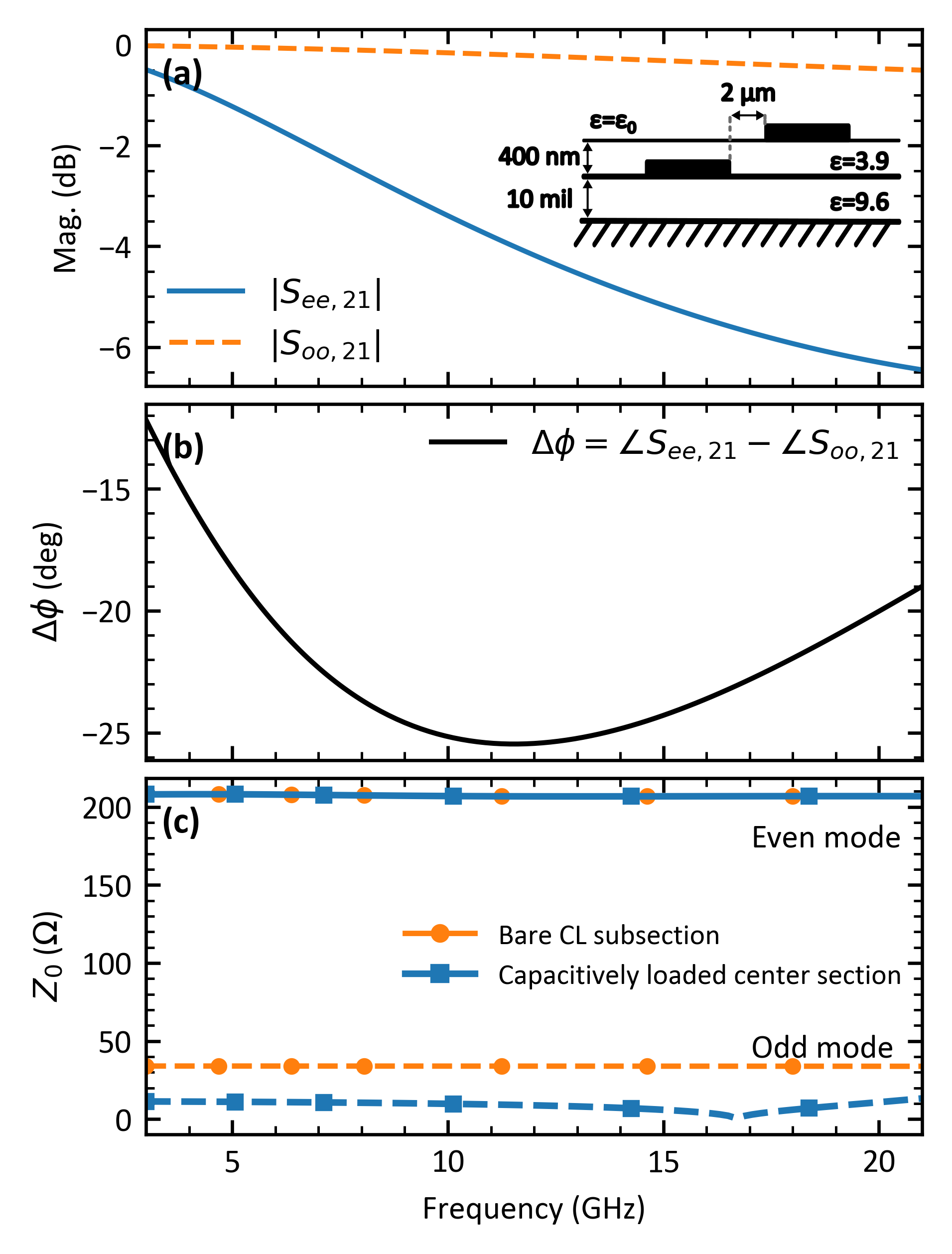}
\caption{Modal response of the bare coupled-line subsection and comparison with the cascaded model of the center section. The inset shows the cross section of the coupled-line subsection. (a) Magnitudes of the modal transmission coefficients $|S_{ee,21}|$ and $|S_{oo,21}|$ extracted from the EM simulation of the bare coupled-line subsection. (b) Modal phase difference $\Delta\phi=\angle S_{ee,21}-\angle S_{oo,21}$. \corr{(c) Modal impedances $Z_{0e}$ and $Z_{0o}$ of the bare coupled-line subsection, together with the effective modal impedance magnitudes of the capacitively loaded center section obtained from the cascaded model comprising four coupled-line subsections and the physical interline capacitors $C_1=28.35$~fF, $C_2=153.44$~fF, and $C_3=543.90$~fF. The capacitive loading primarily reduces the odd-mode impedance, and leaves the even-mode impedance largely unchanged.}
}
\label{Fig:Model_CL}
\end{figure}

Choosing 15 $\mu m$ line widths with a 2 $\mu m$ gap gives a geometry compatible with the available microfrabrication process. To further ease fabrication, the two coupled conductors are implemented on separate metal layers, laterally offset and isolated by a SiO$_2$ dielectric (see inset in Fig.~\ref{Fig:Model_CL}). This also enables the realization of the MIM loading capacitors without additional vias. Since no simple closed-form model is available for this inhomogeneous multiconductor structure \cite{wei_multiconductor_1984}, the unloaded coupled-line subsection was first analyzed using FEM simulations. The resulting per-unit-length inductance and capacitance matrices were then extracted following the multiconductor transmission-line formalism of \cite{hill_crosstalk_1994, cooke_s-parameter_1992}.
\begin{equation}
\mathbf{L}'=
\begin{bmatrix}
L_{11} & L_{12}\\
L_{12} & L_{11}
\end{bmatrix},
\qquad
\mathbf{C}'=
\begin{bmatrix}
C_{11} & -C_{12}\\
-C_{12} & C_{11}
\end{bmatrix}.
\label{eq:LCmatrices}
\end{equation}

the corresponding even- and odd-mode per-unit-length parameters are
\begin{equation}
L'_e=L_{11}+L_{12}, \qquad L'_o=L_{11}-L_{12},
\label{eq:Lmodes}
\end{equation}
\begin{equation}
C'_e=C_{11}+C_{12}, \qquad C'_o=C_{11}-C_{12}.
\label{eq:Cmodes}
\end{equation}
From these quantities, the modal characteristic impedances and propagation constants of the bare coupled-line subsection are
\begin{equation}
Z_{0e,\mathrm{CL}}=\sqrt{\frac{L'_e}{C'_e}}, \qquad
Z_{0o,\mathrm{CL}}=\sqrt{\frac{L'_o}{C'_o}},
\label{eq:Z0bare}
\end{equation}
\begin{equation}
\gamma_e=j\omega\sqrt{L'_eC'_e}, \qquad
\gamma_o=j\omega\sqrt{L'_oC'_o}.
\label{eq:gammabare}
\end{equation}

Table~\ref{tab:center_bare_modal} summarizes the extracted modal parameters of the unloaded coupled-line subsection for the chosen manufacturable line widths and gap. The corresponding modal transmission magnitudes and phase difference are shown in Fig.~\ref{Fig:Model_CL}(a) and Fig.~\ref{Fig:Model_CL}(b), respectively. \corr{The resulting modal impedances $Z_{0e}$ and $Z_{0o}$ (Fig.\ref{Fig:Model_CL}c)} show that the even-mode impedance is already at the desired value, while the odd-mode impedance is too high. Additional capacitive loading is therefore required to lower the odd-mode impedance and achieve the required coupling.

\begin{table}[!t]
\caption{Extracted modal parameters of the bare coupled-line subsection at \(f_0=12\) GHz.}
\label{tab:center_bare_modal}
\centering
\begin{tabular}{lcc}
\hline
Parameter & Even mode & Odd mode \\
\hline
\(Z_0\) (\(\Omega\)) & 205.7 & 31.415 \\
\(L'\) (nH/mm) & 1656.0 & 232.7 \\
\(C'\) (pF/mm) & 39.2 & 235.6 \\
\hline
\end{tabular}
\end{table}
Each coupled-line subsection of length \(l_{\mathrm{CL}}\) is then represented, for mode \(m\in\{e,o\}\), by the transmission matrix
\begin{equation}
\mathbf{T}_m^{\mathrm{CL}}=
\begin{bmatrix}
\cosh(\gamma_m l_{\mathrm{CL}}) & Z_{0m,\mathrm{CL}}\sinh(\gamma_m l_{\mathrm{CL}})\\[4pt]
\dfrac{1}{Z_{0m,\mathrm{CL}}}\sinh(\gamma_m l_{\mathrm{CL}}) & \cosh(\gamma_m l_{\mathrm{CL}})
\end{bmatrix}
\label{eq:Tcl}
\end{equation}

The coupling capacitors are introduced between the distributed coupled-line blocks. For the even mode the two conductors are at equal potential and thus the coupling capacitors have no effect. However, for the odd mode the conductors carry opposite voltages and each capacitor contributes an effective shunt admittance twice that of its nominal value:
\begin{equation}
Y_{C_k,o}=j\omega(2C_k), \qquad C_k \in {C_1,C_2,C_3}.
\label{eq:Ycapodd}
\end{equation}
which is represented by the transmission matrix
\begin{equation}
\mathbf{T}_{C_k,o}=
\begin{bmatrix}
1 & 0\\
j\omega(2C_k) & 1
\end{bmatrix}.
\label{eq:Tcapodd}
\end{equation}

The full center section is therefore obtained by cascading the corresponding modal matrices. For the even mode,
\begin{equation}
\mathbf{T}_e=
\mathbf{T}_e^{\mathrm{CL}_1}
\mathbf{T}_e^{\mathrm{CL}_2}
\mathbf{T}_e^{\mathrm{CL}_3}
\mathbf{T}_e^{\mathrm{CL}_4}
\label{eq:Te_total}
\end{equation}
and for the odd mode,
\begin{equation}
\mathbf{T}_o=
\mathbf{T}^{C_1}_{o}\mathbf{T}_o^{\mathrm{CL}_1}
\mathbf{T}^{C_2}_{o}\mathbf{T}_o^{\mathrm{CL}_2}
\mathbf{T}^{C_3}_{o}\mathbf{T}_o^{\mathrm{CL}_3}
\mathbf{T}^{C_2}_{o}\mathbf{T}_o^{\mathrm{CL}_4}
\mathbf{T}^{C_1}_{o}
\label{eq:To_total}
\end{equation}

For a symmetric reciprocal section, the effective modal characteristic impedance can be obtained from the resulting ABCD matrix as
\begin{equation}
Z_{0m}=\sqrt{\frac{B_m}{C_m}},
\qquad m\in\{e,o\}.
\label{eq:Z0eff}
\end{equation}

The coupling capacitances were then numerically optimized to minimize the deviation of both the coupling factor and characteristic impedance from their ideal values, yielding:
\[
Z_{0e}=207.0\,\Omega,
\qquad
Z_{0o}=10.3\,\Omega.
\]
for coupling capacitor values of:
\[ C_1 = 28.35~\mathrm{fF}\qquad C_2 = 153.44~\mathrm{fF}\qquad C_3 = 543.90~\mathrm{fF}\]
The resulting effective modal impedances of the full center section are shown in Fig.~\ref{Fig:Model_CL}(c). These capacitance values were as an initial estimate for the full-wave optimization. The complete five-section hybrid, including the loaded center section, bends, discontinuities, MIM capacitor geometry, and port transitions, was optimized as a single structure using FEM simulations. \corr{The SiO$_2$ dielectric layer was then modeled with a loss tangent of $\tan\delta = 1\times10^{-3}$.}

\section{Fabrication}

\begin{figure}
\centering
\includegraphics[width=3.4in]{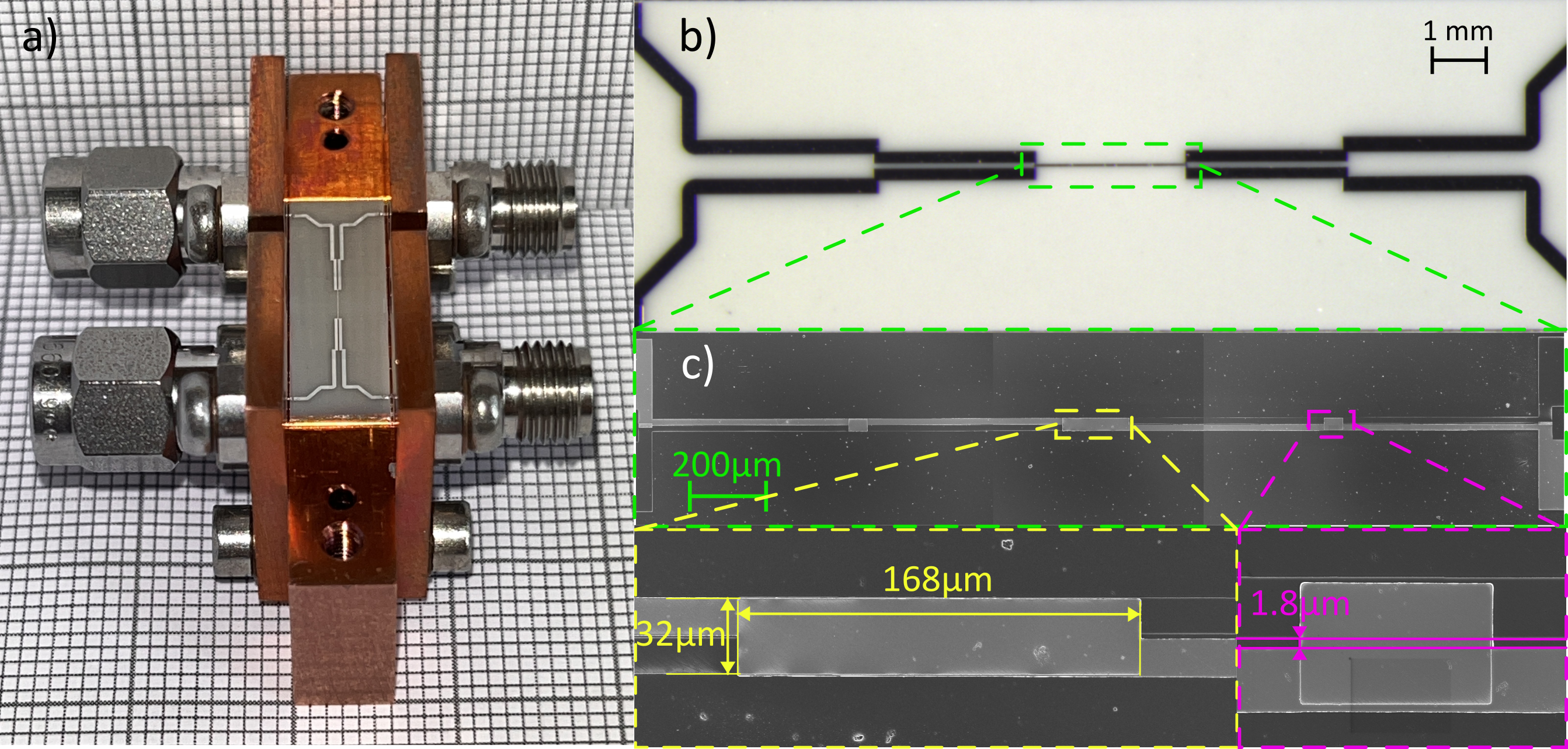}
\caption{(a) External view of the fabricated 4--20~GHz quadrature hybrid. The dimensions excluding connectors are $6.0 \times 40.0 \times 23.0$~mm$^3$ (X$\times$Y$\times$Z). (b) Optical micrograph of the hybrid chip showing the alumina substrate (white) and the two niobium branch microstrip lines. (c) SEM micrograph of the central section of the hybrid. The inset shows a zoom-in of the broadside capacitive coupling, highlighting the central capacitor and one of the intermediate capacitors.}
\label{Image}
\end{figure}

The hybrid is implemented in superconducting niobium microstrip on a $254\,\mu$m alumina substrate ($\varepsilon_r=9.6$), with a $350$\,nm SiO$_2$ dielectric layer used in the multilayer coupling regions (Fig.~\ref{Image}c). Alumina provides a stable dielectric constant over a wide temperature range and excellent mechanical robustness under cryogenic cool-down, making it well suited for superconducting IF circuitry. The overall chip size is $16.8 \times 6$\,mm$^2$. Sections~1, 2, 4, and~5 are realized as conventional edge-coupled $\lambda/4$ microstrip sections.

A 300\,nm thick niobium base layer was deposited by magnetron sputtering and patterned to form the first (bottom) branch of the hybrid. Photolithography was performed using a DWL2000 laser writer (Heidelberg Instruments) with S1813 photoresist. The pattern was transferred into the Nb layer by inductively coupled plasma reactive ion etching (ICP-RIE) using a CH$_4$/O$_2$ gas mixture.

\corr{After definition of the first Nb layer, the structure was planarized with SiO$_2$ before deposition of the interlayer dielectric.} A 350~nm thick SiO$_2$ dielectric layer was then deposited by magnetron sputtering to form the interlayer dielectric for the side-coupled section and capacitors. The second (top) branch of the hybrid was then patterned using the same laser writer. Precise alignment between the lower and upper conductors, better than 200~nm, was required in the central edge-coupled region to achieve the designed capacitive coupling. Following development, a second Nb layer was sputtered and defined by lift-off.

\corr{Openings in the sputtered SiO$_2$ dielectric for the contact pads are defined by lift-off during the SiO$_2$ deposition. The Pd contact pads are subsequently deposited followed by a lift-off step}. (see Fig.~\ref{fig_0}). The samples were diced and mounted onto a copper carrier using silver epoxy. 

RF interfacing was achieved using 2.92-mm K-type connectors. The transition from coaxial connectors to microstrip lines was implemented using Anritsu K-100 hermetic glass bead feedthroughs with sliding contacts soldered directly to the microstrip pads \cite{meledin20191mm}.

\section{Cryogenic Performance}
\subsection{Four-Port Characterization}
Characterization of microwave components at cryogenic temperature is non-trivial due to the presence of long stainless-steel coaxial cables between the 300\,K and 4\,K stages. These cables introduce temperature-dependent attenuation and phase shifts that can't be accurately removed using a simple room-temperature calibration. Therefore, careful de-embedding of the cryogenic interconnects is required.

The measurement configuration follows the methodology reported in~\cite{rashid_superconducting_2014}.  
A room-temperature four-port calibration was first performed at the cryostat reference plane using the UOSM method, with unknown-through, open, short, and matched-load standards. Inside the cryostat, the DUT reference plane is connected via cryogenic coaxial cables experiencing a strong temperature gradient upon cooling down.

To account for the temperature-dependent loss and phase changes of the cryogenic interconnects, a four-port short-circuit calibration fixture was mounted at the DUT reference plane and measured at 4\,K. The resulting cryogenic short measurement was then used for direct fixture compensation of the cables between the room-temperature calibration plane and the DUT plane, following~\cite{rashid_superconducting_2014}. \corr{After compensation, the four cold-short standards exhibited an RMS residual of approximately 0.007--0.008~dB in magnitude and 0.05--0.06$^\circ$ in phase over 4--20,GHz, with maximum deviations of approximately 0.033~dB and 0.29$^\circ$, respectively. Although, this doesn't include residual source/load-match and tracking errors, which aren't fully determined by a single-standard correction. The adopted room-temperature calibration followed by cold short compensation is a compromise for four-port cryogenic characterisation and follows the approach previously used for superconducting IF hybrids~\cite{rashid_superconducting_2014}. Though, residual mismatch errors can produce standing-wave ripples, especially toward the upper end of the frequency band.}

\begin{figure}[!t]
\centering
\includegraphics[width=3.4in]{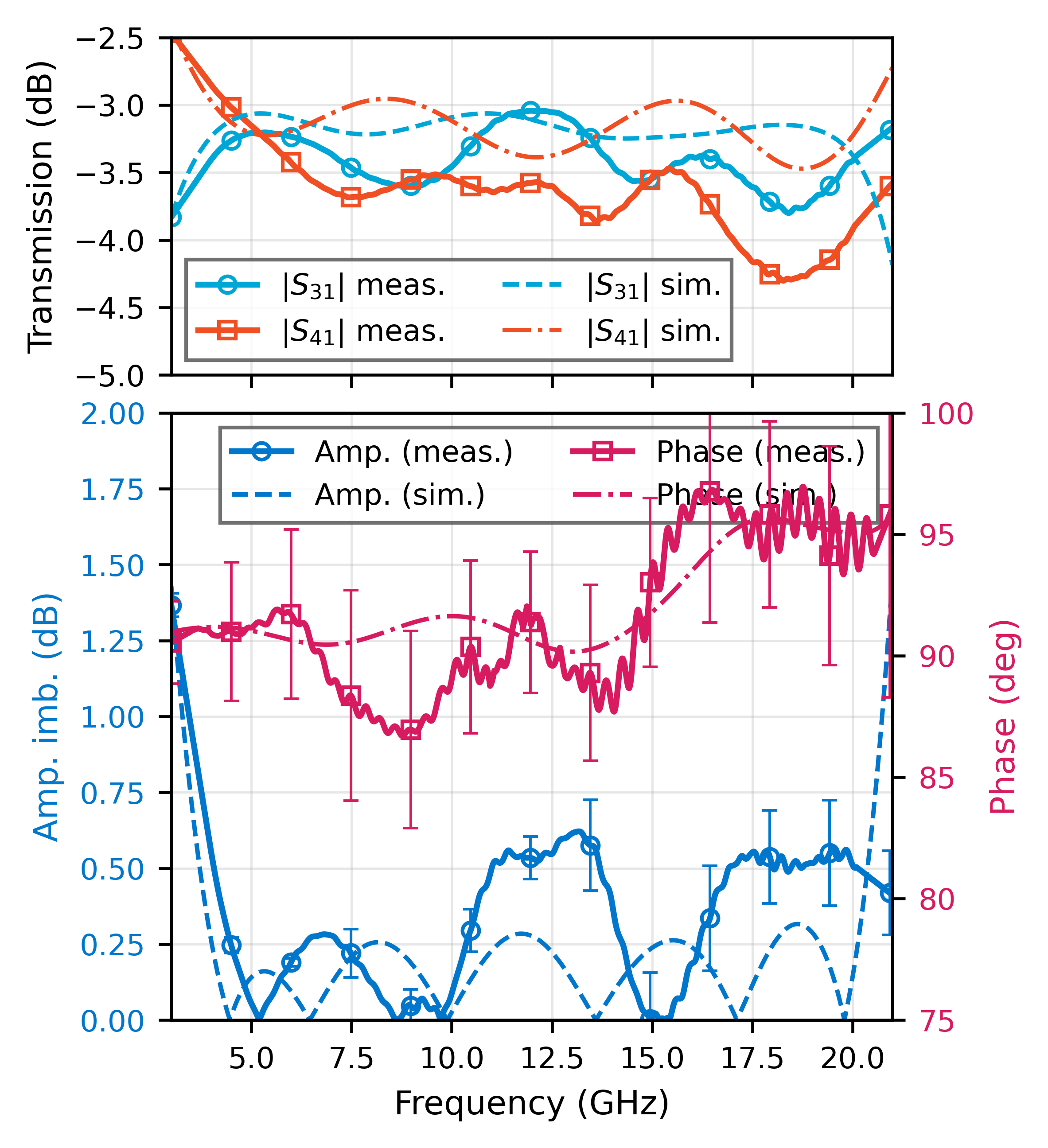}
\caption{Measured and simulated performance of the 4--20\,GHz superconducting IF hybrid. (a) Transmission coefficients $|S_{31}|$ and $|S_{41}|$. (b) Corresponding amplitude imbalance and phase imbalance between the through and coupled ports. \corr{The error bars represent the standard deviation between the imbalance values independently calculated from the equivalent port combinations of the measured four-port S-parameters.}}
\label{fig_2}
\end{figure}

\begin{figure}[!t]
\centering
\includegraphics[width=3.4in]{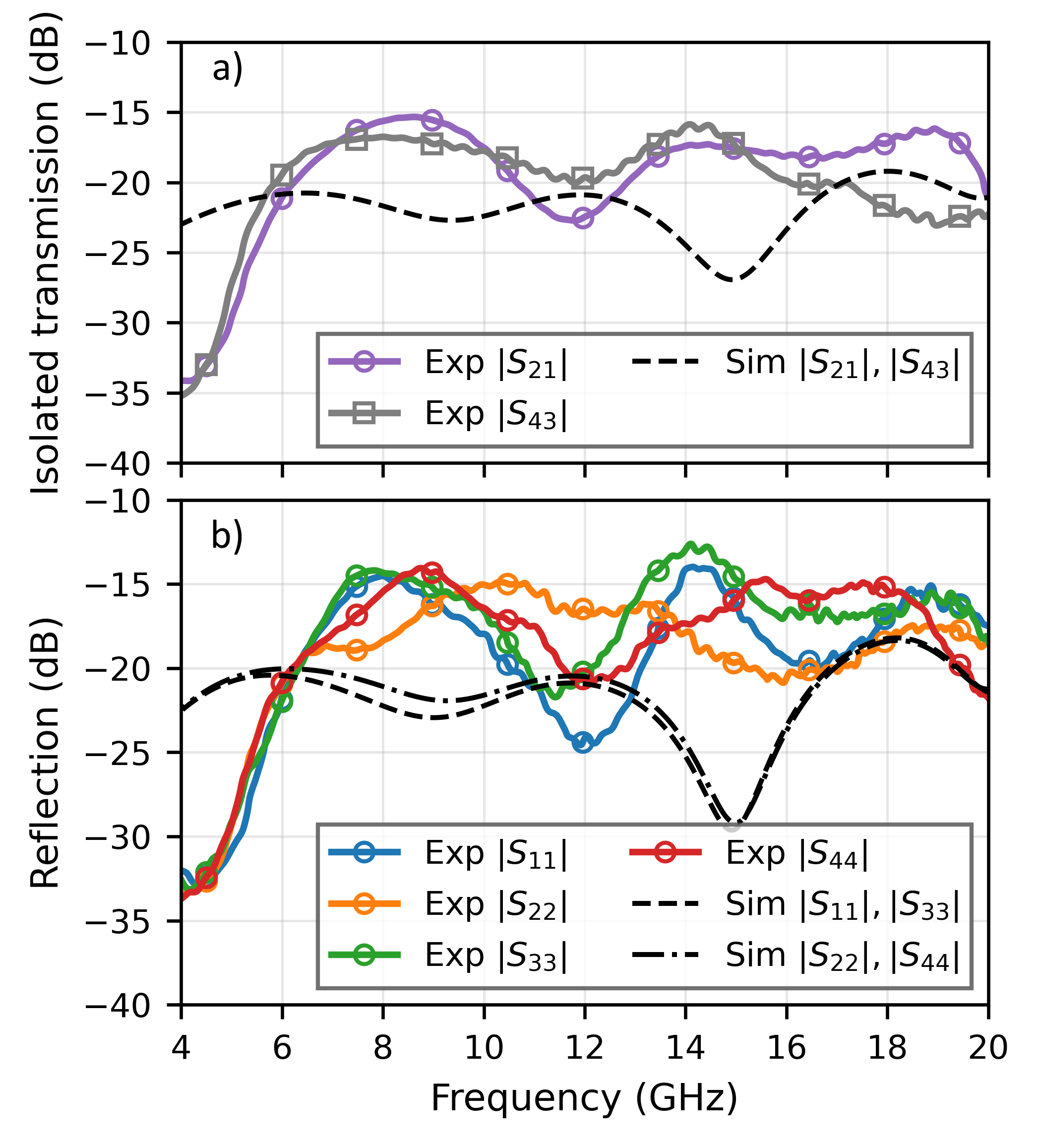}
\caption{Isolation and reflection characteristics of the 4--20\,GHz hybrid. (a) Measured and simulated isolation ($|S_{21}|$ and $|S_{43}|$). (b) Return loss at all four ports.}
\label{fig_3}
\end{figure}

After applying this compensation, the DUT S-parameters were measured at 4\,K. All four S-parameters were measured simultaneously to ensure proper port terminations and accurate extraction of transmission, isolation, amplitude imbalance,
\begin{equation}
\Delta A = 20\log_{10}\left|\frac{S_{21}}{S_{31}}\right|,
\end{equation}
and phase imbalance,
\begin{equation}
\Delta \phi = \angle S_{21} - \angle S_{31} - 90^\circ,
\end{equation}
across the full 4--20\,GHz IF band. \corr{The single-standard short/open compensation corrects the frequency-dependent transmission loss and recovers the electrical reference plane, but it doesn't correct the reflection terms of the cryogenic measurement. Some residual mismatch remains and appears as standing-wave ripple in the through parameters, becoming more pronounced toward the upper end of the band as the measurement uncertainty increases.}
Figs.~\ref{fig_2} and~\ref{fig_3} summarize the measured cryogenic performance of the hybrid. For clarity, the transmission and imbalance are shown for excitation at port~1 only, since the device is symmetric and the remaining ports exhibit comparable behavior.

Fig.~\ref{fig_2} shows the measured and simulated transmission coefficients. Although the overall response follows the expected broadband quadrature behavior, the measured transmission deviates from the ideal simulated hybrid, particularly above 15\,GHz where an increased insertion loss is observed. In addition, the frequency at which the through and coupled branches intersect differs slightly from the simulated response, and the two branches rise together at the upper end of the band rather than exhibiting the ideal complementary behavior.
The resulting amplitude and phase imbalance are shown in the lower panel of Fig.~\ref{fig_2}. Despite the deviations in transmission magnitude, the amplitude imbalance remains below 0.6\,dB across the entire 4--20\,GHz IF band, and the phase imbalance remains within $\pm6^\circ$. These values approach the design targets for 2SB operation and provide the principal basis for evaluating the hybrid performance in this work.

Fig.~\ref{fig_3} shows the measured isolated transmission and return loss of all four ports at 4~K. The hybrid exhibits isolation and reflection levels of the order of -15~dB across most of the 4--20~GHz band, which is higher than the simulated performance. The discrepancies between measurement and simulation are attributed to fabrication sensitivity of the capacitive coupling geometry and to residual uncertainties in the cryogenic calibration and fixture compensation. Given the multi-section broadband nature of the design, small dimensional variations can produce measurable shifts in coupling and phase balance, particularly at the upper end of the band. In a 2SB receiver, however, imperfect port matching is not only a secondary standalone metric. Reflections from the RF hybrid, IF hybrid, SIS mixers, loads, and amplifier inputs can create asymmetric interference paths that limit the image rejection, even when the hybrid amplitude and phase balance are good \cite{hesper_high-performance_2017,khudchenko_modular_2017}. The measured return loss and isolation are therefore relevant for the receiver-level IRR, not only for the mismatch loss of the hybrid. The measured isolation is lower than predicted by simulation, but the matching remains suitable for use in the 2SB IF chain. Its impact at receiver level is discussed in the IRR measurements presented below.

\subsection{Image Rejection Ratio}
The ultimate and most relevant validation of the IF hybrid is obtained at the receiver level by measuring the sideband separation (image rejection) in a complete 2SB system~\cite{rashid_superconducting_2014}. 
Accordingly, the hybrid was integrated in a 2SB. The receiver \cite{lapkin2025Dev, lapkin2026Dev} comprises an RF quadrature hybrid feeding two nominally identical mixer chips~\cite{pavolotsky_fabrication_2026}, followed by a broadband 20-to-50~$\Omega$ IF impedance transformer and the superconducting hybrid reported here. \corr{Each hybrid output is subsequently amplified by a cryogenic Yebes HEMT low-noise amplifier \cite{lopez-fernandez_16-ghz_2024} and two room-temperature Low Noise Factory amplifiers, providing approximately 60~dB of total IF gain over the 4--20~GHz band.}

\corr{The receiver was characterised at an LO frequency of 320~GHz using an automated measurement setup comprising a single LO source}, a pilot tone for sideband measurements, and broadband hot/cold calibration loads for noise temperature and gain calibration~\cite{strandberg_setup_2024, joint_3d-structured_2026}. 

\begin{figure}[!t]
\centering
\includegraphics[width=3.4in]{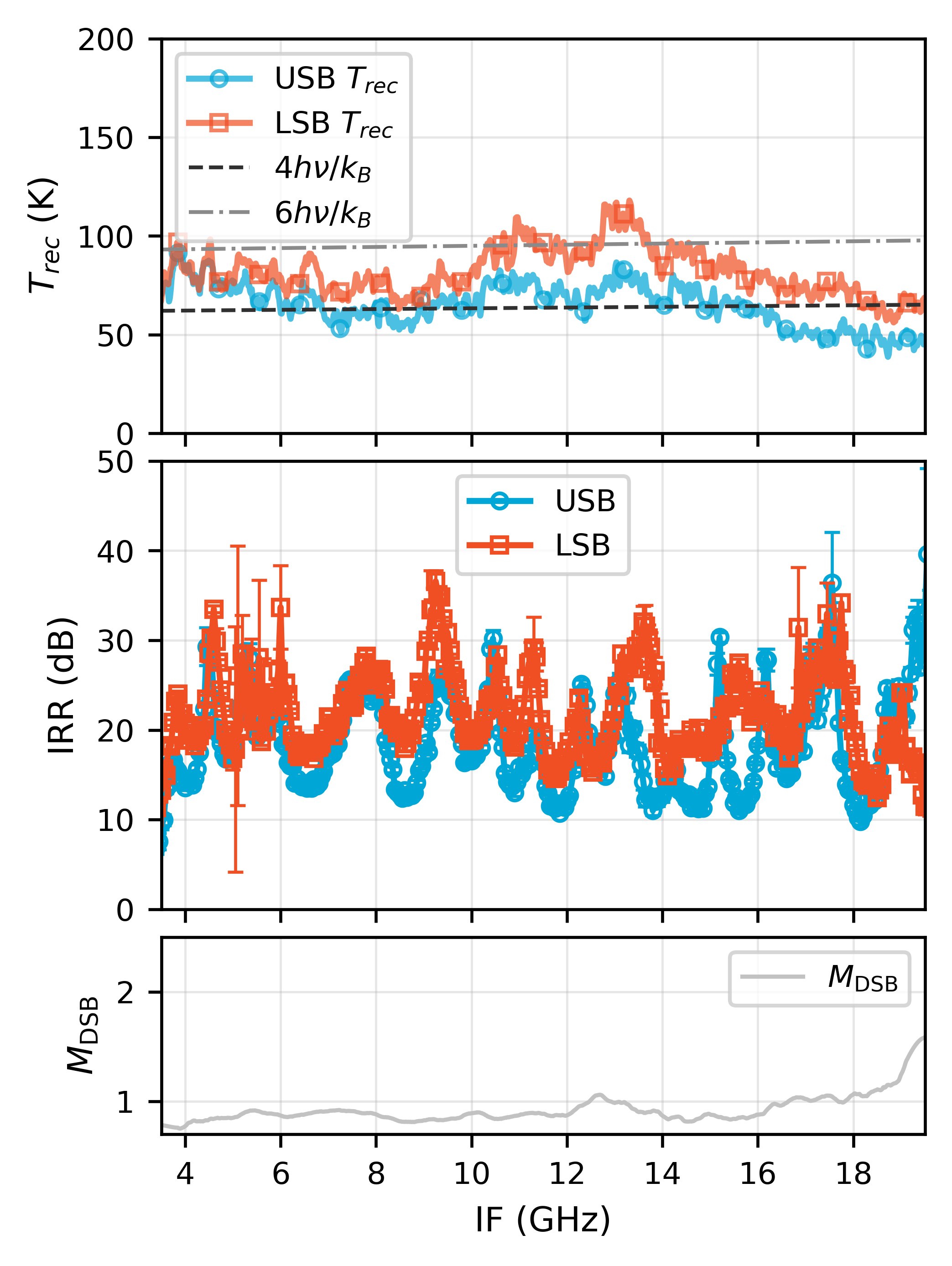}
\caption{System-level performance of the 2SB receiver using the 4--20\,GHz superconducting IF hybrid at $f_{\mathrm{LO}} = 320$\,GHz. 
(a) Measured single-sideband noise temperature versus IF frequency for the upper sideband (USB) and lower sideband (LSB). 
(b) Corresponding image rejection ratio (IRR) extracted using the Kerr method for USB and LSB.  \corr{Error bars show the propagated IRR uncertainty from the integrated CW-tone powers and the $M_{\mathrm{DSB}}$ correction.} (C) Amplitude imbalance, \corr{$M_{DSB}$}, between the two signal paths in linear scale.}
\label{fig_4}
\end{figure}

The image rejection ratio (IRR) was determined following the method described by Kerr \textit{et al.} in ALMA Memo 357~\cite{kerr2001alma}. 
\corr{In this approach, three measured power ratios are used. The quantities $M_U=(P_1/P_2)_U$ and $M_L=(P_1/P_2)_L$ are obtained by injecting a continuous-wave (CW) test tone in the upper and lower sidebands, respectively, and $M_{\mathrm{DSB}}=\Delta P_1/\Delta P_2$ is obtained from the IF power differences measured using broadband hot (295~K) and cold (77~K) calibration loads.} The CW test tones doesn't require calibrated RF power levels, since only relative IF responses are used. 
From these measured quantities, the image rejection ratios at the two IF ports are derived as:
\begin{equation}
R_1 = \frac{M_U}{M_L} \cdot \frac{M_L - M_{\mathrm{DSB}}}{M_U - M_{\mathrm{DSB}}}, 
\end{equation}
\begin{equation}
R_2 = \frac{M_L}{M_U} \cdot \frac{M_U - M_{\mathrm{DSB}}}{M_L - M_{\mathrm{DSB}}},
\end{equation}
\corr{Here, $R_1$ and $R_2$ denote the image rejection ratios measured at the two outputs of the IF hybrid, corresponding to the two sideband-separated receiver output channels.
}
This method provides an accurate determination of the IRR even when the relative RF test-signal amplitudes are unknown, and remains valid for wideband 2SB receivers. The resulting IRR values across the 4--20~GHz IF band are shown in Fig.~\ref{fig_4} (middle panel), together with the corresponding single-sideband noise temperatures (top panel) and the measured IF-output power ratio $M_{DSB}$ (bottom panel). $M_{DSB}$ is used in the IRR extraction and represents the measured power ratio between the two IF outputs. It therefore includes not only the IF-hybrid imbalance, but also amplitude differences arising from unequal conversion gain, cryogenic IF gain, and room-temperature amplifier gain.

\corr{As seen in Fig.~\ref{fig_4}a, the USB and LSB receiver-noise temperatures follow each other closely across the full 4--20~GHz IF band and exhibit the same overall spectral features. The LSB noise temperature is higher by 10~K on average, indicating a small residual asymmetry between the two receiver branches; unequal coupling of LO amplitude noise may contribute to this offset. The value of $M_{DSB}$ remains typically between approximately 0.85 and 1.05 over most of the band, indicating that the two receiver branches are reasonably well balanced in gain (see Fig.~\ref{fig_4}c).}

\corr{The receiver IRR was measured at 50~MHz intervals across the IF band, giving an average rejection values of about 18.5~dB for the USB and 21.9~dB for the LSB over IF band. The rejection remains  above 11~dB for the USB and 15~dB for the LSB over nearly the entire band. The frequency-dependent structure of the IRR is consistent with residual mismatch and reflection-induced interference within the complete receiver signal path, as previously observed in 2SB SIS receivers~\cite{khudchenko_design_2023}. Despite these receiver-level effects, the measurement demonstrates sideband separation over the full wideband IF range and therefore validates the operation of the proposed hybrid at the system level.}

\begin{table}[t]
\centering
\caption{Comparison with representative analog 2SB SIS receivers. 
IRR values refer to receiver-level measurements.}
\label{tab:irr_comparison}
\scriptsize
\setlength{\tabcolsep}{2.5pt}
\begin{tabular}{lccc}
\hline
Reference & IF (GHz) & IF hybrid & IRR (dB) \\
\hline
Rashid et al.~\cite{rashid_superconducting_2014}
& 4--8
& SC thin-film
& $>15$ \\

Khudchenko et al.~\cite{khudchenko_modular_2017}
& 4--12
& Cryogenic
& $>11.5$ \\

Hesper et al.~\cite{hesper_high-performance_2017}
& 4--12
& Cryogenic
& $>15$ \\

Hesper et al.~\cite{hesper_alma_2022}
& 4--20
& Cryogenic
& $>10$ (19.0/20.97 avg.)$^{a}$ \\

This work
& 4--20
& SC thin-film
& $>10$ (18.5/21.9 avg.) \\
\hline
\multicolumn{4}{l}{\footnotesize $^{a}$Mostly over the practical 4--18~GHz range.}
\end{tabular}
\end{table}

\corr{A comparison with representative analog 2SB SIS receiver demonstrations is given in Table~\ref{tab:irr_comparison}. Earlier superconducting and cryogenic 2SB implementations typically operated over 4--8~GHz or 4--12~GHz IF bands, with receiver-level IRR values of approximately 10--15~dB or higher. More recent wideband receiver have also demonstrated operation extending toward 4--20~GHz \cite{hesper_alma_2022}, although with some degradation of the usable rejection bandwidth near the band edges. The present demonstrates that a planar superconducting thin-film quadrature hybrid, directly characterized at cryogenic temperature over 4--20~GHz, can support receiver-level image rejection above 10~dB across this complete IF range. Although, direct comparison of reported IRR values should be made with care, since the measured rejection also depends on the receiver architecture, characterization method, detection backend, measurement bandwidth, and treatment of the residual noise floor, particularly when the rejected-sideband signal approaches the measurement sensitivity. Within this context, the results establish the proposed hybrid as a suitable broadband IF building block for future wideband 2SB receiver systems.}

\section{Conclusion}
We have designed, fabricated, and cryogenically characterized a 4--20\,GHz superconducting IF quadrature hybrid intended for wideband sideband-separating receivers in radio astronomy. Standalone four-port measurements at 4\,K show amplitude imbalance below 0.6\,dB and phase imbalance within $\pm6^{\circ}$ across the full IF band. The hybrid was further validated in a 2SB receiver configuration, where the measured image rejection exceeds 10\,dB over the full IF band. Although the receiver-level IRR reflects the cumulative imbalance of the complete 2SB chain (including the RF hybrid and mixer asymmetries), these results are consistent with the standalone hybrid characterization and confirm that the IF hybrid does not dominate the observed IRR.

The five-section architecture based on capacitively loaded edge coupled transmission-line sections enables multi-octave operation without the increased design and fabrication complexity of Lange implementations at very large bandwidth. The use of superconducting microstrip technology allows close integration with the mixer IF circuitry and supports low-loss cryogenic operation, making the proposed hybrid a practical building block for future wide-IF 2SB receiver upgrades.

\section{Acknowledgments}
This project has received funding from the EU Horizon Europe research and innovation programme under grant agreement No 101 093 934. The authors gratefully acknowledge the support received from the staff at the Swedish national facility for radio astronomy, the Onsala Space Observatory.

\bibliographystyle{IEEEtran}
\bibliography{references, ref2}

\end{document}